# Observation of non-reciprocal wave propagation in a dynamic phononic lattice

*Yifan Wang [1,+], Behrooz Yousefzadeh [1,+], Hui Chen [2], Hussein Nassar [2], Guoliang Huang [2], Chiara Daraio [1]*

[1] Division of Engineering and Applied Science, California Institute of Technology, Pasadena, CA, 91125, USA

[2] Department of Mechanical and Aerospace Engineering, University of Missouri, Columbia, MO, 65211, USA

[+] Y.W. and B.Y. contributed equally to this work.

**Abstract**

Acoustic waves in a linear time-invariant medium are generally reciprocal, however, reciprocity can break down in a time-variant system. In this paper, we report on an experimental demonstration of non-reciprocity in a dynamic one-dimensional phononic crystal, where the local elastic properties are dependent on time. The system consists of an array of repelling magnets, and the on-site elastic potentials of the constitutive elements are modulated by an array of electromagnets. The modulation in time breaks time-reversal symmetry and opens a directional bandgap in the dispersion relation. A theoretical explanation of the observed non-reciprocal behavior is provided as well. This work provides a prototype for developing acoustic diode that can serve in acoustic circuits for rectification applications .

Phononic crystals and metamaterials control acoustic waves through the geometry of their building blocks, engineered with periodic impedance mismatches and/or local resonances [1-7]. The majority of current realizations focus on designing metamaterials in their spatial dimensions, while the material properties remain unchanged over time. This design framework restricts the application of metamaterials in scenarios where material's tunability and adaptivity is required [8,9]. More importantly, in these time-invariant metamaterials, reciprocity holds as a fundamental principle in wave propagation, requiring the transmission of information or energy between any two points in space to be symmetric for opposite propagating directions [10].

However, non-reciprocal materials or devices, i.e., diodes, are usually required for rectification and control of the associated energy flow. Unlike electric diodes, mechanical or acoustic diodes are just starting to be explored [11-18]. Achieveing non-reciprocity in mechanical systems through intrinsic time-reversal symmetry breaking has been demonstrated in strongly nonlinear networks [11,13,14], selective acoustic circulators [15], and topological mechanical insulators [16-18]. In nonlinear systems, the non-reciprocal behavior is a function of the nonlinear potential and may be tuned by the wave amplitude [19,20]. Recently, theoretical proposals [21-24] suggested the use of external, spatio-temporal modulation of material's properties as a mean to achieve non-reciprocity within the linear operating regime.

Here we demonstrate realization of a dynamic phononic lattice in which the elastic properties can vary over time with spatiotemporal modulation. This time dependence leads to novel wave propagation behaviors such as non-reciprocity [21-24], which is very difficult to achieve in time-invariant systems. Though we focus on elastic waves in a magnetically coupled lattice, the concept extends to other types of waves such as thermal diodes [25] and photonic systems [26]. For instance, non-reciprocal propagation in photonic systems was observed in coupled, modulated waveguides [27] where modulation leads to irreversible mode conversion between the two waveguides. As for our system, it behaves as a mechanical diode operating at tunable frequency ranges. Such device may serve in acoustic circuits, like circulators, transducers and imaging systems to rectify mechanical or acoustic energy flows [11].

Experimental realizations of modulation-induced non-reciprocity in a single phononic waveguide require (i) a dynamic lattice with controllable elastic properties, and (ii) a dynamic modulation with speed comparable to the wave propagation velocity. We meet these requirements by building a mass-spring chain of repelling magnets modulated by externally driven coils. The chain consists of 12 ring magnets ($m = 9.8$ g) free to slide on a supporting smooth cylindrical rail as shown in Fig. 1a. The first and last magnets are fixed to the rail (fixed boundary conditions). To dynamically modulate the chain, we introduce electrical coils around the 8 central ring magnets (masses 3 to 10). The electrical coils are positioned coaxially with the magnets and rest at the same center positions $x_{0,n}$ as shown in Fig. 1a. When a current flows through the electrical coils, they create local magnetic fields that couple to the ring magnets. When the ring magnets are at rest ($x_{0,n}$ position), they sit at the apex of the magnetic potential created by the coils and their coupling forces vanish. When the ring magnets displace, they experience either restoring or repelling forces from the coils, depending on the current direction. The coupling between each pair of ring magnet and coil is similar to a grounding spring. When the grounding spring stiffness is modulated spatiotemporally, time-reversal symmetry is broken leading to the formation of a non-reciprocal bandgap in the dispersion diagram [21-24] as shown in Fig. 1b.

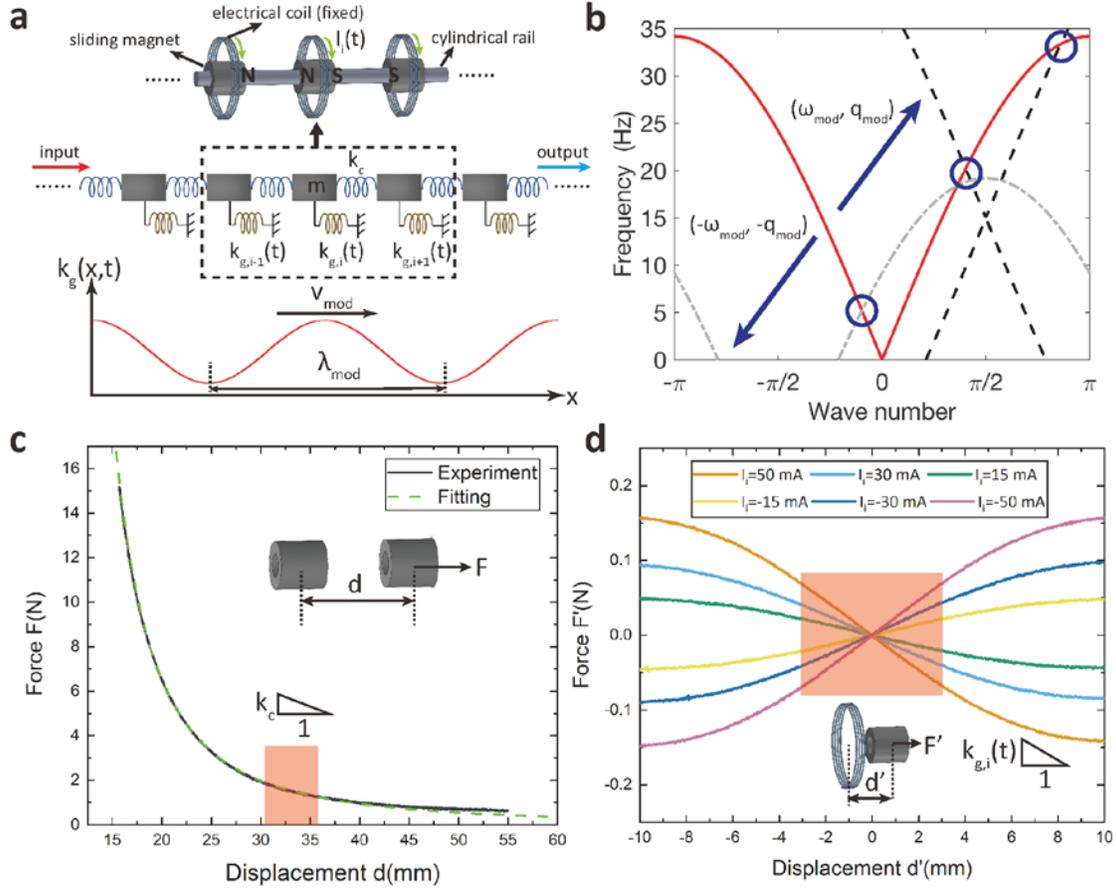

**Figure 1** Experimental setup for the non-reciprocal dynamic phononic lattice. (a) Top: Schematic of the experimental setup. Middle: Discrete mechanical representation of the system with masses and springs. Bottom: Schematic illustration of the modulation concept by changing the gounding spring stiffness ($k_g$) in a wave-like fashion. (b) Scattering analysis: The red solid curve describes the original dispersion relation of the un-modulated monatomic lattice. The black dashed and grey dash-dotted curves correspond to Floquet-Bloch replicas of the original curves obtained by translation along the solid blue arrows $\pm(\omega_{mod}, q_{mod}) = \pm(15\text{Hz}, \pi/2)$. Parity-breaking crossings (circled) are where Bragg's condition is satisfied and non-reciprocal wave scattering is anticipated. (c) Force-displacement curve for neighboring magnetic masses, measurement (solid) and fitted curve (dashed). (d) Measured force-dispacement curve between the ring magnet and its surrounding coil at different currents. The red shaded regions in both (c) and (d) corresponds to the dynamic operating regime of our experiments.

To characterize the mechanical parameters of our system, we measure the repelling force between neighboring masses as a function of their displacement (see Supplemental Material). The resulting force-displacement curve exhibits a nonlinear force that is characteristic of dipole repulsion shown in Fig. 1c. We also measure the force between the magnets and the surrounding coils at different applied currents in Fig. 1d. To measure the dynamic response of the system, we drive the 2nd mass with a sinusoidal force of frequency $f_{dr}$, and the velocity of mass 11 is monitored with a laser vibrometer (output signal). The velocity response is measured using a lock-in amplifier as a function of different $f_{dr}$ for different modulation parameters. Due to the small vibration amplitude of the driving signal ($\leq$ 5 mm), the coupling between masses can be approximated by a linear response in the red shaded area of Fig. 1c. The linearized coupling stiffness between adjacent magnets obtained from experiments is $k_c \approx 113$ N/m. Similarly the coupling between the electromagnets and the masses can be linearized in the dynamic regime of interest in Fig. 1d. We consider only nearest neighbor interactions between masses and mass-coil pairs, since non-nearest neighbour interactions decay to a negligible amount (see Supplemental Material).

The spatiotemporal modulation of the system can be achieved by applying sinusoidal AC currents through the coils. Each coil is subjected to a current of the same frequency, $f_{mod}$, but with a phase shift of $\pi/2$ or $-\pi/2$ between neighbours. The equivalent grounding stiffness for the $n$-th mass thus can be modelled as:

$$k_{g,n} = k_{g,DC} + k_{g,AC} \cos\left(2\pi f_{mod} t \mp \frac{\pi x_{0,n}}{2a}\right) = k_{g,DC} + k_{g,AC} \cos(2\pi f_{mod} t \mp q_{mod} n) \quad (1)$$

where $k_{g,DC}$ is the small time-independent grounding stiffness added by the on-site electromagnetic force, $k_{g,AC}$ is the modulation amplitude of the grounding stiffness, $x_{0,n}$ is the equilibrium position of each unit, and $q_{mod} = \pm\pi/2$ is the normalized wave number. Equation (1) describes a traveling wave with wavelength $\lambda_{mod} = 4a$ and speed $v_{mod} = 4a f_{mod}$. The modulation amplitude measured in our experiments is $k_{g,AC} = 24$ N/m, which is 21% of the coupling stiffness, $k_c$. The constant part of the grounding stiffness is $k_{g,DC} = 2.4$ N/m, which is one order of magnitude smaller than the oscillatory component.

In the absence of modulation ($k_{g,AC} = 0$), the dispersion relation for an incident small-amplitude plane wave $u_0(n,t) = U_0 \exp(i(qn - \omega t))$ is described by $D(\omega, q) = k_{g,DC} - m\omega^2 + 4k_c \sin^2\left(\frac{q}{2}\right) = 0$. Modulating the lattice harmonically with ($f_{mod}, q_{mod}$) generates an additional scattered field $u_s(n,t) = U_s \exp(i(q_s n - \omega_s t))$ whose mode is *shifted* by an amount ($\omega_{mod}, q_{mod}$) due to spatiotemporal periodicity: ($\omega_s, q_s$) = ($\omega_0, q_0$) ± ($\omega_{mod}, q_{mod}$). The scattered field is negligible however ($U_s \ll U_0$) except when it is resonant with the incident field; i.e., when the modified Bragg's condition $D(\omega_s, q_s) = D(\omega_0, q_0) = 0$ is met [22]. Graphically, scattered modes are located at crossings between the original ($D(\omega, q) = 0$) and shifted ($D(\omega_s, q_s) = 0$) dispersion curves. Note that the crossings are non-symmetrically distributed in a way that breaks parity of the dispersion diagram and, ultimately, reciprocity of wave propagation. Depending on whether $q_0 q_s$ is positive or negative, the scattered mode propagates either with or against the incident wave, i.e., is either transmitted or reflected. In both cases however, its frequency is shifted away from the incident frequency $\omega_0$. This translates into a one-way dip in the transmission spectrum around $\omega_0$.

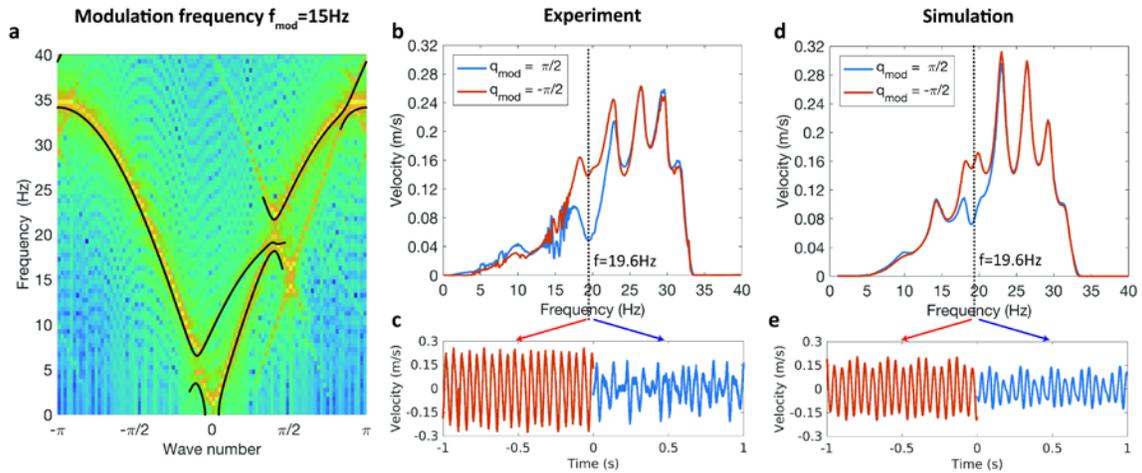

**Figure 2 Non-reciprocal wave propagation for $f_{mod} = 15$ Hz.** (a) Dispersion diagram of the modulated lattice calculated by Fourier analysis of simulated velocity fields (color map)

and analytically by coupled mode theory (solid black line). (b) Measured velocity response function. The amplitude ratio at 19.6 Hz is $r = 2.9$. (c) Measured velocity time series at $f_{dr} = 19.6$ Hz. The time series for $q_{mod} = -\pi/2$ is shown along the negative time axis for better illustration. (d) and (e) are the simulation results corresponding to (b) and (c), respectively. The simulated amplitude ratio at 19.6 Hz is $r = 1.9$ in panel (d).

We first set the modulation frequency to $f_{mod} = 15$ Hz, which falls within the pass band of the monoatomic lattice. For this modulation frequency, three crossings exist at 5 Hz, 19 Hz and 33 Hz and non-reciprocal wave characteristics are anticipated for neighboring driving frequencies $f_{dr}$ as shown in Fig. 2a. We measure the velocity of the last mass in the array as a function of the driving frequency, $f_{dr}$ in Fig. 2b. The velocity profiles differ when the acoustic waves are traveling in the same (red) or opposite (blue) direction to the modulation wave, at driving frequencies close to $f_{dr} = 19.6$ Hz. We define the co-directional/contradirectional bias ratio as $r = U^-/U^+$ where $U^\mp$ denotes the velocity response amplitude for $q_{mod} = \mp\pi/2$. At $f_{dr} = 19.6$ Hz, the measured velocity response profile in time shows that waves traveling in opposite directions have different amplitudes and profiles, with a bias of $r \approx 2.9$, shown in Figs. 2b, c. The time-domain amplitudes are lower than the amplitudes obtained from the velocity response functions. This is due to the anharmonic nature of the response in the modulated lattice. However, results demonstrate that the signal transfer around $f_{dr} = 19.6$ Hz is strongly enhanced when traveling along the modulation direction and suppressed in the other direction, thus exhibiting a non-reciprocal behavior.

We developed a mathematical model to capture the dynamic characteristics of the modulated lattice. The system can be described as:

$$m\ddot{u}_n + F_{loss} + k_{g,n}u_n + F_{coupl} = \delta_{2,n}A\cos(2\pi f_{dr}t) \qquad (2)$$

for $1 \leq n \leq 12$. Here, $u_n(t) = 0$ at the two boundaries $n = 1,12$. $F_{loss} = b\dot{u}_n + \mu\,\text{sign}(\dot{u}_n)$ represents dissipative forces within the chain, with viscous damping coefficient $b = 0.056$ kg/s and Coulomb friction coefficient $\mu = 0.012$ N (see Supplemental Material). The coupling force term is $F_{coupl} = P(a - u_n + u_{n+1}) - P(a - u_{n-1} + u_n)$, where we use the approximation $P(x) = c_1/(x - c_2)^2$ with $c_1 = 0.9788$ mNm² and $c_2 = 7.748$ mm obtained from a fitting based on Fig. 1c. $\delta_{2,n}$ is the Kronecker delta which is 1 for $n = 2$ and zero everywhere else. The forcing amplitude $A = 0.21$ N is obtained as a fitting parameter. At this value of the forcing amplitude, the response of the system is well approximated by the linearized equations of motion (the contribution from nonlinearity is discussed in the Supplemental Material). The experimental and numerical velocity response functions for a non-modulated lattice agree well [28] (see Supplemental Material). When the modulation is turned on, the velocity profiles obtained in experiments and simulations show a similar nonreciprocal response in Figs. 2d, e. However, the non-reciprocal behavior at $f_{dr} = 19.6$ Hz is less pronounced in simulations than in measurements ($r \approx 1.9$).

We computed dispersion curves from space-time Fourier analysis of the velocity field and compared them with the ones obtained with the plane-wave expansion method in Fig. 2a. The observed non-reciprocal wave characteristics, at $f_{dr} = 19.6$ Hz, agree well with the dispersion characteristics. The dispersion curves in Figs. 1b & 2a predict non-reciprocal behavior also near 5 Hz and 33 Hz. However, the experimental velocities are too small at these frequencies to capture the effect. Note that the analyses (numerical and theoretical) on an infinite lossless lattice (Fig 2a) predicted the same frequency range for non-reciprocal wave propagation as the experiments (Fig 2b) and simulations (Fig 2d) on a finite lossy lattice. The effects of energy loss and finite number of units are therefore secondary to modulation effects; see Supplemental Material for discussions of finite-size and loss effects.

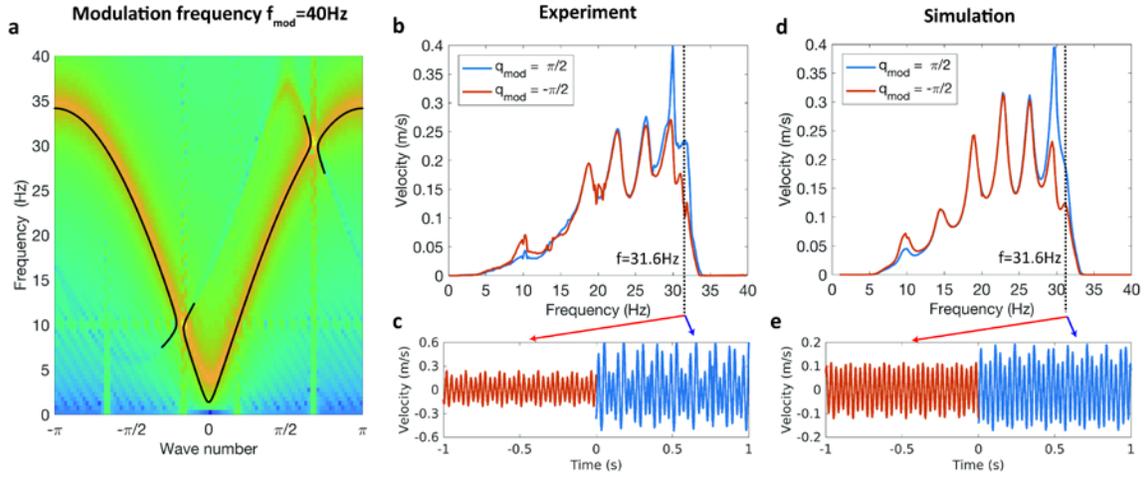

**Figure 3. Non-reciprocal wave propagation for $f_{mod} = 40$ Hz.** (a) Dispersion diagram of the modulated lattice calculated by Fourier analysis of simulated velocity fields (color map) and analytically by coupled mode theory (solid black line). (b) Measured velocity response function. The amplitude ratios are $r = 1.8$ at 9.8 Hz and $r = 0.4$ at 31.6 Hz. (c) Measured velocity time series at $f_{dr} = 31.6$ Hz. The time series for $q_{mod} = -\pi/2$ is shown along the negative time axis for better illustration. (d) and (e) are the simulation results corresponding to (b) and (c), respectively. The simulated bias ratios are $r = 1.6$ at 9.8 Hz and $r = 0.7$ at 31.6 Hz in panel (c).

In order to demonstrate the tunability of the non-reciprocal frequency range in our system, we next set the modulation frequency to $f_{mod} = 40$ Hz, within the band gap of the underlying monatomic lattice. Our model predicts non-reciprocal wave behavior for driving frequencies near the crossings at 10 Hz and 30 Hz as shown in Fig. 3a. This is also captured in the measured velocity responses in Fig. 3b and time domain profiles at $f_{dr} = 31.6$ Hz in Fig. 3c. Corresponding numerical simulations in Figs. 3d,e agree very well with the measurements.

The dispersion curve of the modulated lattice in Fig. 3a obtained from numerical calculation corroborates the observed non-reciprocal characteristics for $f_{mod} = 40$ Hz. It reveals two crossings located near 30 Hz and 10 Hz and visible as small bright yellow regions lying on a main dispersion branch. At these points, the modulation-induced scattered field is strong enough to change the overall wave field. This manifests in the velocity response functions as $r > 1$ near 10 Hz and $r < 1$ near 30 Hz. For other points along the main dispersion branch, the scattered wave is too weak compared to the incident field to induce any noticeable non-reciprocal effects. In contrast to the case for $f_{mod} = 15$ Hz, the crossing here occurs between a positive and a negative branch of the dispersion curve ($\omega_0 \omega_s < 0$) and leads to the opening of a couple of "vertical" bandgaps as shown in Fig. 3a. Such crossings in infinite loss-less systems are characteristic of unstable interactions caused by supersonic modulation velocities, where the velocity field is continuously amplified by drawing energy from the modulation [31,32]. However, our experimental system is intrinsically lossy and finite, and remains stable in the studied regime. The presence of losses is known to quench instabilities [33]. In our system, this translates in the presence of a sharp peak around 30 Hz in the transmission spectrum, shown in Figs. 3b,d.

In conclusion, our results provide an experimental demonstration of modulation-induced non-reciprocity in a linear phononic lattice. The operating range of our lattice is beyond the asymptotic limits that are typically enforced in the existing theoretical work. The experimental realization of dynamically modulated nonreciprocal systems opens new opportunities for sound and vibration insulation [11,12,15], phononic logic [13,14] and energy localization and trapping [34]. In the future, the phononic waveguide developed in our work could be employed to study the nonlinear dynamics of modulated lattices, a regime that has not been explored before. The design could also be miniaturized into micro- or nano-scale electromechanical systems [35-37] with tunable frequencies as basic elements for acoustic rectifying circuits.

**Acknowledgements**

Y.W., B.Y. and C.D. acknowledge the support from the National Science Foundation under EFRI Grant No. 1741565. H.C., H.N. and G.H. acknowledge support from the National Science Foundation under EFRI Grant No. 1641078. B.Y. acknowledges the support from the Natural Science and Engineering Research Council of Canada.

**Author contributions**

Y.W. and C.D. designed the experiment. Y.W. performed the experiments. B.Y. performed analytical and numerical modelling of the system. H.C., H.N. and G.H. performed analytical calculations on the dispersion curves. Y.W., B.Y., H.N., G.H. and C.D. wrote the manuscript. All authors interpreted the results and reviewed the manuscript.

**Competing Financial Interests**

Nothing to report.

# Supplemental Material for 'Observation of non-reciprocal wave propagation in a dynamic phononic lattice'

Yifan Wang [1,+], Behrooz Yousefzadeh [1,+], Hui Chen [2], Hussein Nassar [2], Guoliang Huang [2], Chiara Daraio [1]

[1] Division of Engineering and Applied Science, California Institute of Technology, Pasadena, CA, 91125, USA

[2] Department of Mechanical and Aerospace Engineering, University of Missouri, Columbia, MO, 65211, USA

[+] Y.W. and B.Y. contributed equally to this work.

**Experimental details**

Figure S1 shows a photograph of the experimental setup. The ring magnets used in our experiment are NdFeB Grade N42 magnets with dimensions of 12.7mm OD, 6.4 mm ID and 12.7mm in length (K&J Magnetics, Inc.). The cylindrical rail is made of fiberglass with diameter 4.76mm. Sleeve bearings made from PTFE with 6.4mm OD and 4.8mm ID and are installed in each ring magnets to fit between magnets and rail and reduce sliding friction. The ring magnets are placed on the rail with the same polarization facing each other, causing them to repel and form a regular one-dimensional lattice. By fixing the two end magnets, the mass chain reaches equilibrium with a uniform spacing $a = 33.4$ mm between neighbours ($a$ is the lattice constant). Electrical coils (APW Company) used in this experiment have dimensions 48.3 mm OD, 27.1 mm ID and 17.5 mm in length, with inductance of 104 mH. To measure the force-displacement interaction between two ring magnets and between a magnet and a coil with flowing current, the magnet or coil is fixed on two testing plates of an Instron E3000 materials tester with a 250N load cell. The magnets and/or coil are aligned coaxially and force-displacement curves are recorded for over 3 times and averaged to reduce noise in the data.

To achieve dynamic modulation on the current flowing in the 8 electrical coils, we use electrical signals generated by two synchronized function generators A and B (Agilent 33220A). A phase shift of $\pm\pi/2$ is set between two function generators for forward and backward modulations.

The electrical coils #1 and #5 (phase offset 0) are connected in parallel to function generator A, while coils #3 and #7 (phase offset $\pm\pi$) are conneted to function generator A but with reversed polarization to achieve the π phase shift. Similarly, coils #2 and #6 (phase offset $\pm\pi/2$) are connected parallel to function generator B, while #4 and #8 (phase offset $\pm 3\pi/2$) are connetcted to B with reversed polarization. The two function generators are set to the same frequency $f_{mod}$ and amplitude, while the phase shift is set to either $\pi/2$ or $-\pi/2$ for backward and forward modulation directions. The modulation current in each coil is checked with an oscilloscope to ensure same amplitude and phase lag of $\pi/2$ between neighbors. Due to the small cross-inductance and low operation frequency, we do not observe current induced from cross-inductance between neighbors in these coils.

The phononic chain is driven by a separate coil placed off-center from the first moving magnet which is connected to a sweeping-frequency lock-in amplifier (Stanford Research SR860). The velocity of the last moving magnet is measured with a laser vibrometer (Polytec CLV-2534).

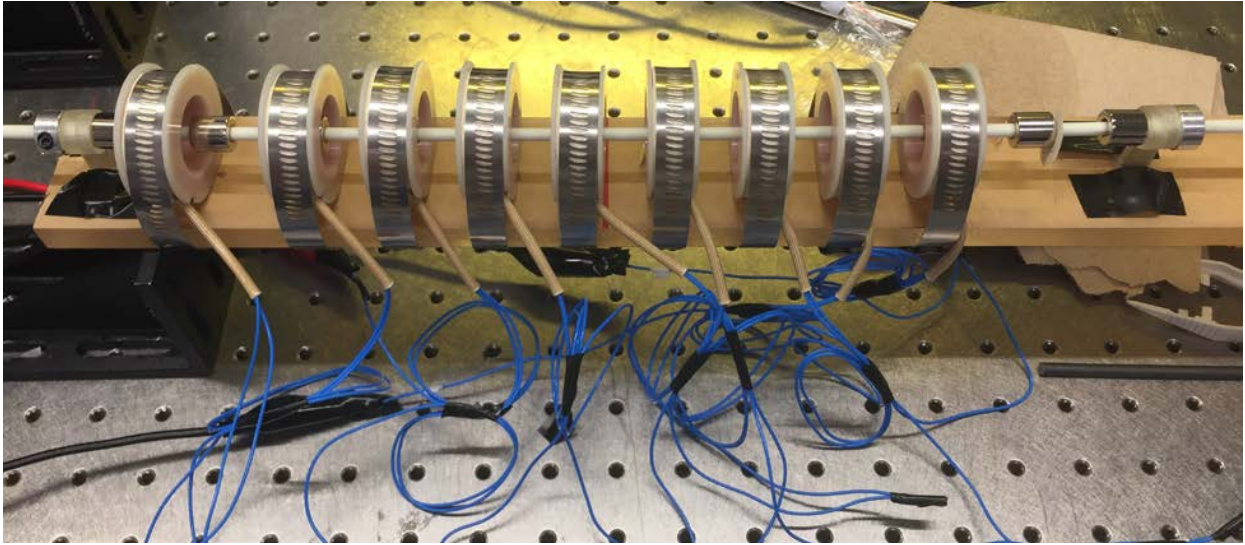

**Figure S1** A photograph of the experimental setup. The center 8 coils are used for dynamic modulation and the left most coil is used for exciting the system.

**Simulation of the velocity response curves**

The control parameters for experiments are the modulation amplitude $k_{g,AC}$, modulation frequency $f_{mod}$ and driving frequency $f_{dr}$. The forcing amplitude $A$ may also be controlled by changing the driving current, but it is kept constant in our experiments. Figure S2 shows the measured and simulated velocity response functions of the un-modulated lattice ($k_{g,AC} = 0$). There is generally very good agreement between measurements and simulations, except for the peak around 15 Hz that is not captured in measurements. This is most likely because the friction forces in experiments are not fully captured by the Coulomb friction model used in simulations. The peaks in the velocity response functions are due to the finite size of the system (waves reflecting from the boundaries) and may be attributed to different modes of vibration – see also the section 'Spatial profile of the steady response' in Supplemental Material. We observe a sharp cut-off around 33 Hz, which is lower than the cut-off value of $(1/\pi)\sqrt{k_c/m} = 34.3$ Hz based on the dispersion relation. This is due to the presence of energy dissipation [29].

To simulate the velocity response functions for a given set of control parameters, Equation (2) is solved in time until the initial transients decay. For each $f_{dr}$, the amplitude of motion is then obtained based on the Fourier transform of velocity time series over 500 driving cycles. This procedure reproduces the velocity amplitude measured by the lock-in amplifier. The response of the modulated system is not periodic due to the presence of two incommensurate frequencies $f_{dr}$ and $f_{dr} \pm f_{mod}$ (see, for example, Figures 2b and 2d).

We use the velocity response function of the lattice with no modulation ($k_{g,AC} = 0$) to obtain the coefficients of viscous damping $b$ and Coulomb friction $\mu$, as well as the forcing amplitude $A$. We note that including Coulomb friction is essential for capturing the sharp decay of the velocity response function near the cut off. We used a smooth approximation of the sign function in simulations, $\text{sign}(\dot{u}_n) \approx \tanh(\alpha \dot{u}_n)$ with $\alpha = 1000$.

**Non-nearest neighbor interactions**

To measure the contributions from non-nearest neighbor interactions, the longer-range magnet-magnet and magnet-coil force-displacement curves are measured (Figure S2). Based on the magnet-magnet measurement, the effective spring constant created by the 2$^{nd}$ nearest neighbor (blue box in Figure S2a) is 4 N/m, about 3.5% of the 1$^{st}$ nearest neighbor coupling $k_c = 113$ N/m, which is negligible in our analysis. The magnet-coil interaction measurement shows that the 2$^{nd}$ nearest neighbor magnet-coil coupling spring constant (blue box in Figure S2b) is ~1 N/m, about 4% of our modulation amplitude of $k_{g,AC} = 24$ N/m and less than 1% of the coupling stiffness $k_c = 113$ N/m, which is negligible as second order term. Based on these measurements, we only consider the nearest neighbor interactions between magnets and electrical coils.

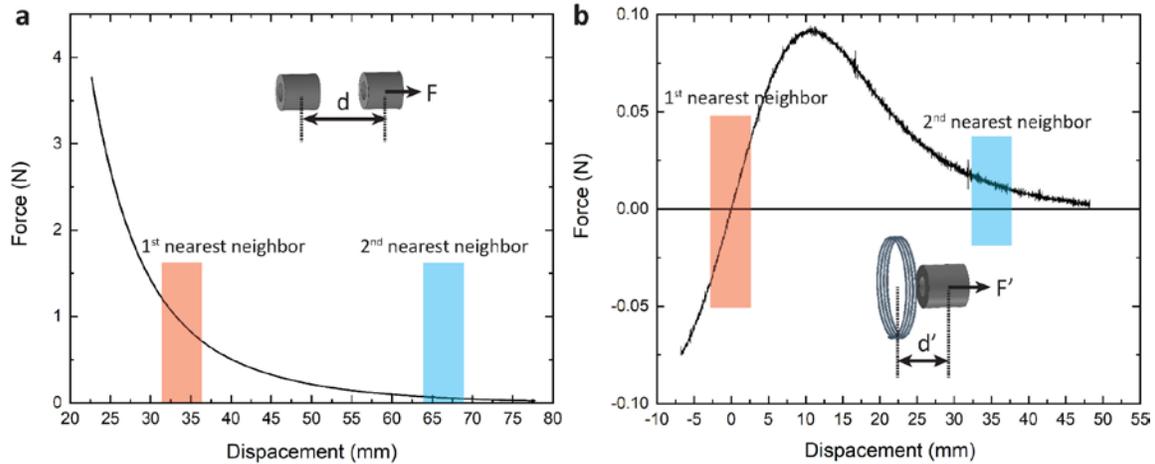

**Figure S2** Comparison between 1$^{st}$ and 2$^{nd}$ nearest neighbor interactions for (a) magnet-magnet repulsion and (b) magnet-coil interaction at current of 30 mA. The red and blue shaded boxes correspond to the operating regions of the 1$^{st}$ and 2$^{nd}$ nearest neighbors.

**Contribution from nonlinearity**

To assess the importance of nonlinearity in the coupling force, $F_{coupl}$, we computed the velocity response function for two cases: once with $F_{coupl}$ as described for Equation (2) and again with a linear approximation, $F_{coupl} \approx k_c(2u_n - u_{n-1} - u_{n+1})$. Figure S3 compares these simulated velocity response functions to measurements. We see that the resonant peaks shift to higher frequencies due to the hardening nature of the nonlinear force. We used the nonlinear coupling force in all the other simulations in this work because it leads to better agreement with measurements. The nonlinear behavior is most pronounced near the resonance peaks due to higher amplitudes of motion. However, it is important to note that the influence of nonlinearity is very weak: using a linearized coupling force results in less than 2.2% error in the locations of the resonant peaks in velocity response functions of Figure S3. This is why we could use the linear theory (e.g. dispersion curves) to predict and explain where non-reciprocity occurs.

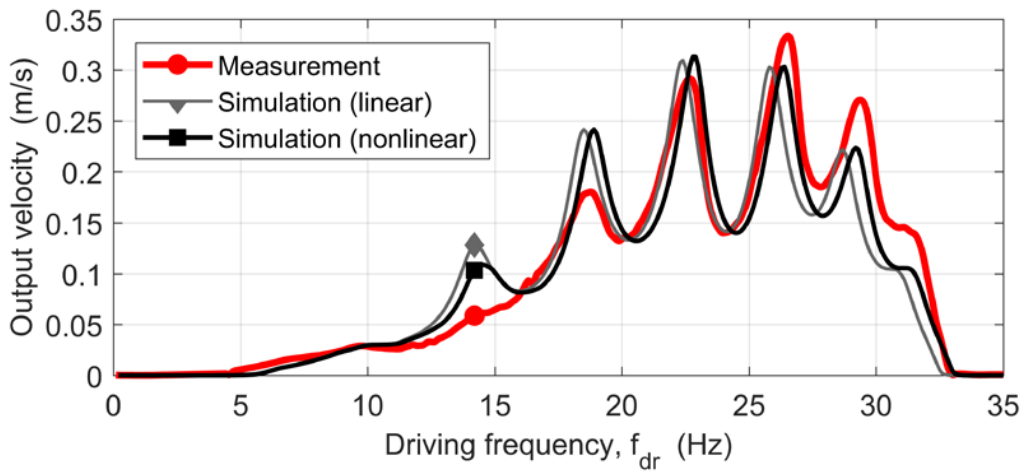

**Figure S3** Velocity response curves of the unmodulated lattice ($k_{g,AC} = 0$). The thick red curve (circle marker) denotes the measured response. The grey curve (diamond marker) and black curve (square marker) denote simulated responses using a linear and a nonlinear coupling force, respectively.

**Calculation of dispersion curves**

Dispersion relations describe the (free) propagation of plane waves through a medium. We computed the dispersion curves of the modulated lattice using two different methods.

The spectral energy density method [21] was used first to obtain the dispersion relation based on direct numerical simulation of the linearized loss-less lattice with $N = 128$. We applied an impulse (initial velocity) to the middle unit and computed the transient response until the impulse reached the boundaries. We then obtained the space-time Fourier transform of the velocity field. The highest contour plot of the resulting field in the frequency-wavenumber space gives the dispersion diagram of the lattice. This is a common method for direct calculation of the dispersion curve for modulated lattices in the literature [21-23].

In the second method, we adopted a plane wave expansion of the displacement field $u_n(t)$ of the form $\sum_j U_j \exp i(q_j n - \omega_j t)$ with $(\omega_j = \omega + j\omega_{mod}, q_j = q + jq_{mod})$. Upon substitution in the linearized motion equation of an ideal infinite lattice, namely

$$m\ddot{u}_n(t) = -k_{g,n}(t)u_n(t) + k_c \left(u_{n+1}(t) + u_{n-1}(t) - 2u_n(t)\right) \qquad (S1)$$

it comes that

$$\begin{pmatrix} D(\omega_{-1}, q_{-1}) & -1 & 0 \\ -1 & D(\omega_0, q_0) & -1 \\ 0 & -1 & D(\omega_1, q_1) \end{pmatrix} \begin{pmatrix} U_{-1} \\ U_0 \\ U_1 \end{pmatrix} = \begin{pmatrix} 0 \\ 0 \\ 0 \end{pmatrix} \qquad (S2)$$

where the expansion of the displacement was truncated and only the terms $j = -1, 0, +1$ were kept due to the smallness of $k_{g,AC}/k_c$, and with $D(\omega_j, q_j) = \frac{2\left(m\omega_j^2 - 4k_c \sin^2\left(\frac{q_j}{2}\right)\right)}{k_{g,AC}}$. The dispersion relation is then deduced from the zero-determinant condition

$$D(\omega_{-1}, q_{-1})D(\omega_0, q_0)D(\omega_1, q_1) - D(\omega_{-1}, q_{-1}) - D(\omega_1, q_1) = 0. \qquad (S3)$$

Note finally that the curve generated in this manner is only valid in the vicinity of the unperturbed dispersion curve of the non-modulated lattice given by $D(\omega, q) = 0$.

**Energy loss effects**

Losses are present in the experimental system studied in this paper. In our model, we describe the losses as viscous damping and Coulomb friction (Eq. 2). The dissipation parameters are extracted from experiments as fitting parameters, matching the velocity response of non-modulated systems. We numerically study the effects of energy loss in our systems in two different scenarios: (*I*) pulse propagation and (*II*) continuous driving of the lattice. We use the finite lattice of Eq. (2) with modulation frequency $f_{mod} = 15$ Hz for all the simulations in this section. Similar results are obtained for other modulation frequencies.

(*I*) For pulse propagation, a Dirac impulse is applied to the first unit (as initial velocity) and the response at the end unit is monitored for both forward- and backward-traveling modulation waves. Fig. S4 shows the results of these simulations for three sets of damping and friction: (i) left column: no loss ($b = 0$, $\mu = 0$); (ii) middle column: only viscous damping ($b = 0.056$, $\mu = 0$); (iii) right column: damping and friction ($b = 0.056$, $\mu = 0.003$). The top row (a) in Fig. S4 shows the impulse response of the end unit in time domain, the middle row (b) shows the impulse response in the frequency domain (i.e., the Fourier transform of the top row), and the bottom row (c) shows the response of the entire lattice in the time domain. The time series for backward-traveling waves are shown along the negative time axes for better illustration.

The left column in Fig. S4 shows the impulse response of the end unit in the absence of energy loss. The non-reciprocal behavior around 5 Hz and below 20 Hz is clearly observed in the frequency domain (panel b). We can also observe that the time series for the forward and backward cases are different (panels a and c). As expected, this is consistent with analytical and numerical predictions of the infinite lossless lattice in Fig. 2a. As viscous damping is increased to the value obtained from experiments (middle column in Fig. S4), less energy is transferred to the end unit for both values of $q_{mod}$. This is observed most clearly in the time series (top and bottom rows), but is also visible in the frequency domain (middle row). Most importantly, the non-reciprocal wave propagation around 5 Hz and below 20 Hz persists. When the friction coefficient is increased to $\mu = 0.003$ (right column in Fig. S4), we observe that most of the input energy is lost and little energy is transferred to the end unit. Notably, the transferred energy retains its non-reciprocal nature, consistent with the previous cases. If we repeat the above numerical analysis with the same damping and friction coefficients as in the experiments ($b = 0.056$, $\mu = 0.012$), no energy reaches the end unit. This explains why we were not able to measure the impulse response of the end unit in the experimental setup.

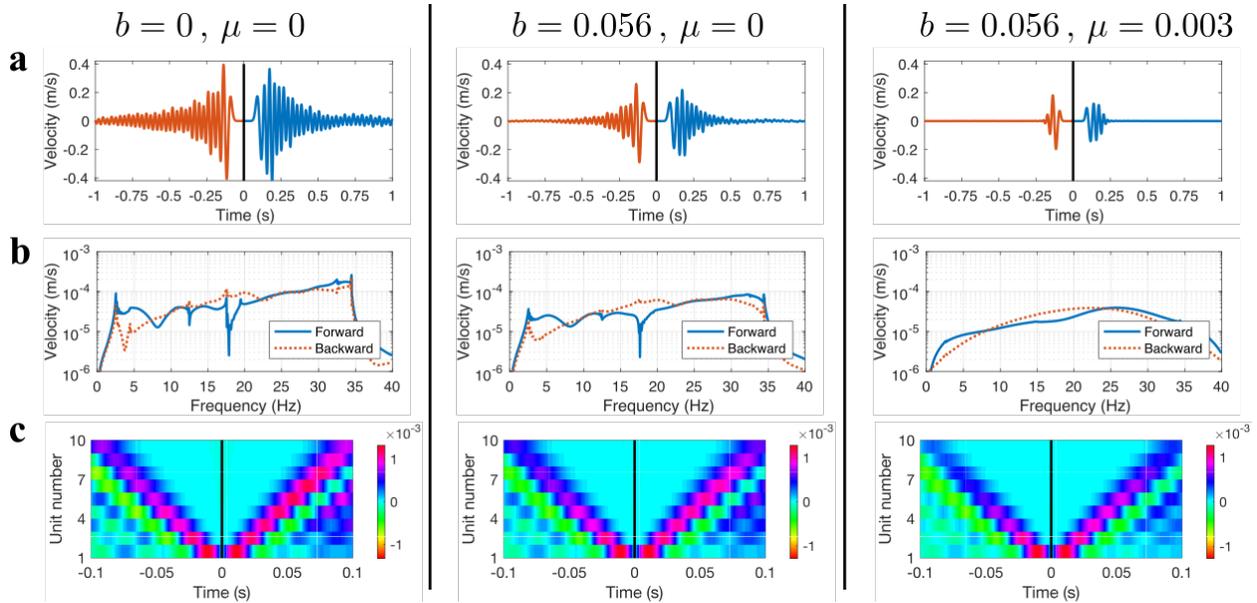

**Figure S4** Influence of energy dissipation on the impulse response of the finite lattice for $q_{mod} = \pi/2$ (forward) and $q_{mod} = -\pi/2$ (backward) at modulation frequency of $f_{mod} = 15$ Hz. Left column: no loss ($b = 0$, $\mu = 0$); (ii) middle column: only viscous damping ($b = 0.056$, $\mu = 0$); (iii) right column: damping and friction ($b = 0.056$, $\mu = 0.003$). The top row (a) shows the impulse response of the end unit in the time domain, the middle row (b) shows the impulse response of the end unit in the frequency domain, the bottom row (c) shows the response of the entire lattice in the time domain. The time series for backward-traveling waves are shown along the negative time axes for better illustration.

(*II*) To evaluate the effect of energy loss for continuous excitation of the lattice, we consider the response of the finite lossy system described by Eq. (2) at modulation frequency $f_{mod} = 15$ Hz. Figure S5 shows the velocity response curves for three different values of energy loss: (a) $b = 0.056$, $\mu = 0.021$ (the same as in the experimental system); (b) $b = 0.056$, $\mu = 0$;

and (c) $b = 0.028$, $\mu = 0$. The results are plotted in logarithmic scale to emphasize the differences between them. Comparing Figs. S5 a & b, we observe that friction plays an important role in suppressing the response at lower frequency where the velocities are small. This explains why the non-reciprocal behavior near 5 Hz was not detectable in experiments. We also observe that the non-reciprocal propagation below 20 Hz is more pronounced in Fig. S5 b than in Fig. S5 a because there is less energy. By further reducing the damping, we observe in Fig. S5 c that the response amplitudes generally increase and the non-reciprocal behavior near 5 Hz becomes more pronounced. Of course, in the limit of zero energy loss the response becomes unbounded at various frequencies because of the continuous excitation.

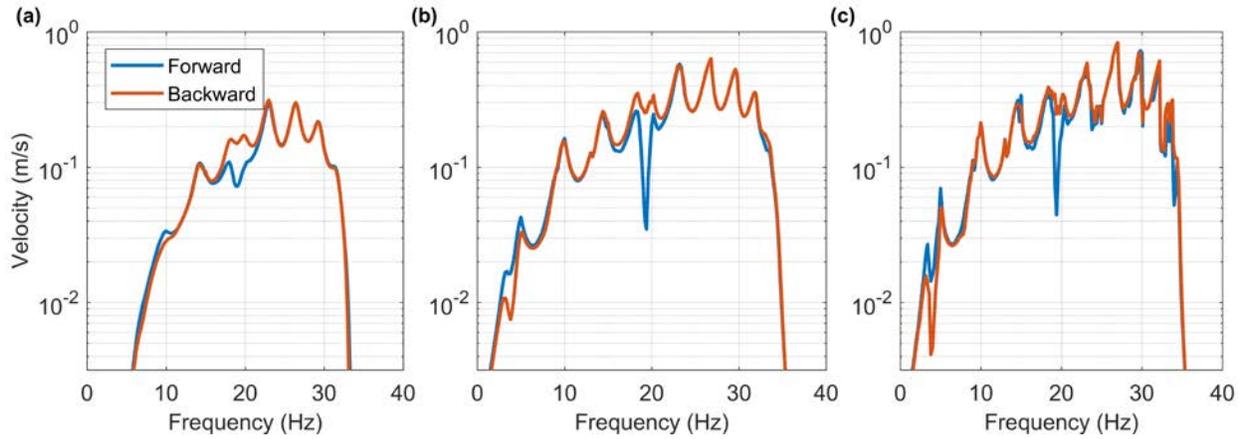

**Figure S5** Velocity response curves at modulation frequency $f_{mod} = 15$ Hz for three different values of energy loss: (a) $b = 0.056$, $\mu = 0.021$ (same as in Fig. 2d); (b) $b = 0.056$, $\mu = 0$; and (c) $b = 0.028$, $\mu = 0$. As energy loss is gradually decreased from (a) to (c), we observe that the response amplitude increases and the non-reciprocal behavior becomes more pronounced, especially near 5 Hz.

**Finite-size effects**

The theoretical findings pertaining to an infinite, lossless medium were able to reliably predict the frequency ranges at which we measured modulation-induced non-reciprocity in a finite lossy medium (see Figs. 2 and 3). In this section, we focus on finite size effects in a modulated lattice. Specifically, we compute the bias ratio $r$ (the ratio of the forward- to backward-traveling waves) as a function of the number of units for the modulation frequency of 15 Hz. To decouple finite-size effects from other effects, we perform all the simulations in this section for an undamped chain, with an absorbing layer attached to the end of the lattice to avoid reflections. To remain consistent with the experimental setup, the first and last units in the lattice are not modulated. An impulse (initial velocity) is applied to the first unit, the governing equation are integrated in time, and the response of the last unit is transformed to the frequency domain for $q_{mod} = \mp \pi/2$ to obtain the bias ratio.

Figure S6 shows the bias ratio, as a function of the lattice size. In each case $N$ denotes the total number of magnets within the lattice. The number of modulated units is $N_{mod} = (N-2)/4$ because each modulated unit comprises 4 magnets. We can see three main regions of biased propagation (modulation-induced non-reciprocity) that persist with increasing number of units: around 5 Hz, around 20 Hz and above 35 Hz. We focus on the region around 20 Hz, bearing in mind that the biased responses around 5 Hz and 20 Hz are due to the same phenomenon. We ignore the region above 35 Hz because it lies within the band gap, where the amplitudes of motion are too small to be of practical relevance.

Figure S6 shows a size-persistent region of biased propagation near 20 Hz, approximately between 17.9 Hz and 22.3 Hz. As expected, increasing the lattice size enhances the bias ratio and the frequency range of biased propagation. We can see that for $N = 18$ (4 modulated units) the biased region is already very similar to the biased region for $N = 410$ (100 modulated units). This frequency range is in agreement with the theoretical predictions of the infinite lattice shown in Fig. 2a. Note that there is a narrow-band unbiased response near 19 Hz for $N \geq 18$. This corresponds to the extra branch of the dispersion curve that penetrates the modulation-induced band gap near 20 Hz, as already depicted in Fig. 2a. This effect is not present in the short lattice because of its narrow-band nature: it is well understood [30] that the dispersion curve of a finite lattice can be obtained by "sampling" the dispersion curve of an infinite lattice at discrete values of wavenumber.

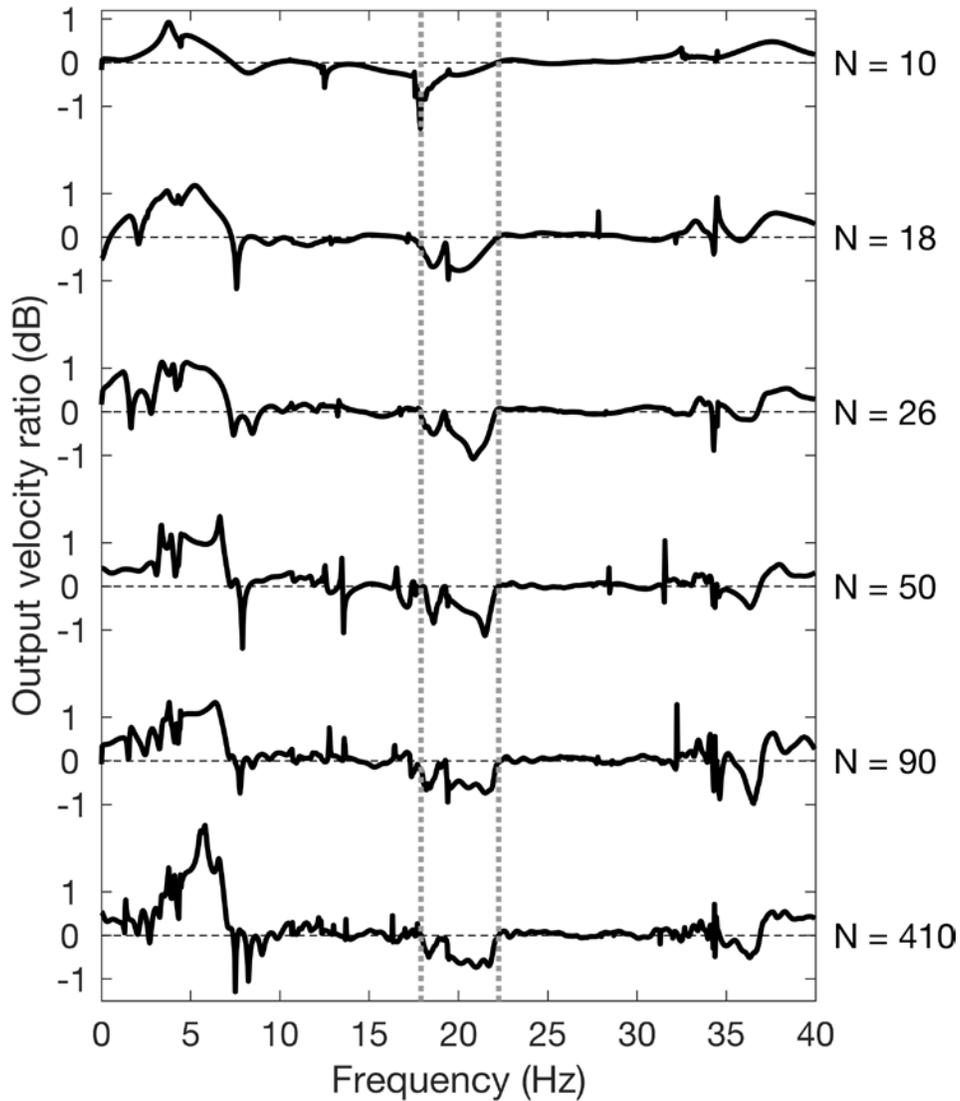

**Figure S6** Ratio of the output velocities of the forward- to backward-traveling waves for modulation frequency $f_{mod} = 15$ Hz as a function of the lattice size. $N$ is the total number of magnets in the lattice. The quantity on the y-axis is $\log(r)$, thus zero corresponds to reciprocal propagation ($r = 1$). The results for different lattice sizes are offset for better illustration. $N =$

10 corresponds to the same number of units as in the experimental setup. The vertical dotted grey lines at 17.9 Hz and 22.3 Hz denote the frequency range for modulation-induced non-reciprocity near 20 Hz for the longest chain.